\documentclass[reprint, aps, prl, amsmath, amssymb]{revtex4-2}

\usepackage{graphicx} 
\usepackage{bm} 
\usepackage{slashed}
\usepackage{tikz}
\usepackage[compat=1.1.0]{tikz-feynman}
\tikzfeynmanset{warn luatex=false}
\usepackage{silence} 
\usepackage{mathtools}
\usepackage{mathrsfs}
\usepackage[colorlinks=true, citecolor=blue!90!black, linkcolor=blue!90!black, linktocpage=true, urlcolor=red!70!black]{hyperref} 

\newcommand{\eg}{\textit{e.g.}}
\newcommand{\ie}{\textit{i.e.}}

\DeclareMathOperator{\sn}{sn}
\DeclareMathOperator{\cn}{cn}
\DeclareMathOperator{\dn}{dn}
\DeclareMathOperator{\Tr}{Tr}
\DeclareMathOperator{\tr}{tr}

\begin{document}

\title{Condensates, Crystals, and Renormalons in the Gross--Neveu Model at Finite Density}

\author{Francesco Benini}
\email{fbenini@sissa.it}
 
\author{Ohad Mamroud}%
\email{omamroud@sissa.it}

\author{Tom\'as Reis}
\email{treis@sissa.it}

\author{Marco Serone}
\email{serone@sissa.it}

\affiliation{SISSA, via Bonomea 265, 34136 Trieste, Italy}
\affiliation{INFN, Sezione di Trieste, via Valerio 2, 34127 Trieste, Italy}

\begin{abstract}

We study the $O(2N)$ symmetric Gross--Neveu model at finite density in the presence of a $U(1)$ chemical potential $h$ for a generic number $a \leq N-2$ of fermion fields. By combining perturbative quantum field theory, semiclassical large $N$, and Bethe ansatz techniques, we show that at finite $N$ two new dynamically generated scales $\Lambda_\mathrm{n}$ and $\Lambda_\mathrm{c}$ appear in the theory, governing the mass gap of neutral and charged fermions, respectively. Above a certain threshold value for $h$, $a$-fermion bound states condense and form an inhomogeneous configuration, which at infinite $N$ is a crystal spontaneously breaking translations. At large $h$, this crystal has mean $\Lambda_\mathrm{n}$ and spatial oscillations of amplitude $\Lambda_\mathrm{c}$. The two scales also control the nonperturbative corrections to the free energy, resolving a puzzle concerning fractional-power renormalons and predicting new ones.

\end{abstract}


\maketitle

In quantum field theory (QFT), strong coupling phenomena --- such as spontaneous symmetry breaking and dynamically generated mass scales --- can often be traced back to nonperturbative effects at weak coupling. Renormalons are a subtle class of such effects, expected to originate from condensates \cite{tHooft:1977xjm, Shifman:1978bx, Parisi:1978az}. In turn, the condensates are proportional to dynamically generated scales, indicating that perturbative techniques can anticipate certain strongly coupled dynamics.

A remarkably fruitful playground to study these effects is the $O(2N)$ Gross--Neveu (GN) model: a theory of $N$ massless Dirac fermions in two spacetime dimensions (2d) coupled via four-fermion interactions \cite{Gross:1974jv}. 
This theory is, notably, a toy version of 4d quantum chromodynamics (QCD) and has diverse applications in condensed matter physics, including conducting polymers \cite{Campbell:1981dc} and inhomogeneous superconductors \cite{Thies:2006ti}.
The model is asymptotically free, and undergoes dynamical mass generation in the infrared (IR) which leads to a gapped phase and spontaneous breaking of a $\mathbb{Z}_2$ chiral symmetry. 
The theory is integrable and its spectrum, which contains kinks and bound states of fermions, as well as its complete $S$\nobreakdash-matrix, are known for any $N$ \cite{Zamolodchikov:1978xm, Karowski:1980kq}.
The tractability of the model prompted a study of the theory at finite density, where the Lagrangian reads
\begin{equation}
\label{eq:UV1}
\mathscr{L} = 
\sum_{j=1}^N \bar\psi^{j} i \slashed{\partial}\psi^{j} 
+ \frac{g^2}2 \biggl( \sum_{j=1}^N \bar\psi^{j} \psi^{j} \!\biggr)^{\!\!2}
+ h \! \sum_{m=1}^a \bar\psi^m \gamma^0 \psi^{m}  \!.
\end{equation}
Here $\psi^j$ are 2d Dirac fermions, $\gamma^\mu$ are 2d gamma matrices, and $h$ is the common $U(1)$ particle-number chemical potential for the first $a \leq N-2$ fermions.

Thanks to a line of research dating back to \cite{Polyakov:1983tt, Forgacs:1991rs}, the relative free energy $\Delta F = F(h)-F(0)$ can be computed with Bethe ansatz techniques. In this way one finds the exact relation between the mass gap $m$ of the theory (at $h=0$) and the dynamically generated scale $\Lambda$
in the $\overline{\rm MS}$ scheme, which at 1-loop is
\begin{equation}
\label{Lambda}
\Lambda \approx h \, e^{ - \tfrac{ \pi }{ (N-1) \, g^2(h) } } \,.
\end{equation} 
More recently, thanks to \cite{Volin:2009wr, Marino:2019eym}, a systematic weak-coupling expansion of the free energy became possible, revealing the resurgent properties of the expansion as well as its connection with renormalons  \cite{Marino:2019eym, DiPietro:2021yxb, Marino:2021dzn, Bajnok:2022xgx}.
If renormalons originate from condensates, the nonperturbative terms should follow the schematic form (up to possible $\log g$ terms) $\sum_{n,k} (\Lambda/h)^{n \Delta} g^{2k}$, where $\Delta$ is the classical dimension of the condensing operator. In the case $a=1$ of the GN model, instead, \cite{Marino:2021dzn} surprisingly found that renormalons appear as powers of $(\Lambda/h)^{(N-1)/(N-2)}$ which does not fit that pattern for any operator.

In this Letter we combine three different techniques --- \ie, perturbative expansion, the semiclassical large $N$ limit, and integrability --- to attain a compelling picture of the IR physics of the Gross--Neveu model for generic values of $a$ and $h$, and confirm the expected relation between renormalons and condensates. We achieve this by identifying in the model two new dynamically generated scales that govern the nonperturbative sectors: 
\begin{equation}
\label{eq:lambda-12}
\begin{aligned}
\Lambda_\mathrm{n} &= 2h \, \Bigl( \frac{\Lambda}{2h} \Bigr)\!\rule{0pt}{1em}^{\frac{N-1}{N-a-1}} \propto h \, e^{ - \tfrac{\pi}{(N-a-1) \, g^2(h) } } \,, \\
\Lambda_\mathrm{c} &= 2 h \, \Bigl( \frac{\Lambda}{2h} \Bigr)\!\rule{0pt}{1em}^{\frac{2(N-1)}{a}} \propto h \, e^{ - \tfrac{2\pi}{ a\, g^2(h) } } \,.
\end{aligned}
\end{equation}
Indeed, we show that the nonperturbative terms in the free-energy transseries
\begin{align}
\label{eq:ts-12}
\Delta F(h) &\sim h^2 \!\! \sum_{m,n\geq 0} \frac{\Lambda_\mathrm{n}^{2m} \Lambda_\mathrm{c}^{2n} }{ h^{2(m+n)}} \, \sum_{\substack{0 \leq k \\ 0 \leq p \leq p_{m,n} \hspace{-2.4em}}} \, a^{[m,n]}_{k,p} \, g^{2k-2k_{m,n}} (\log g)^p \nonumber \\[-.5em]
&\quad - c_0 \Lambda^2 
\end{align}
reproduce the exotic renormalons of \cite{Marino:2021dzn} by identifying them with $\Lambda_\mathrm{n}$, and predict new ones associated with $\Lambda_\mathrm{c}$. Here $k_{m,n}$ and $p_{m,n}$ are nonnegative integers.
Notice that (\ref{eq:ts-12}) is a transseries since for every $m,n,p$ the series is asymptotic in $k$ and only makes sense after Borel resummation (see, \eg, \cite{Sauzin:2014qzt, Aniceto:2018bis, Serone:2024uwz}).
The two new scales arise naturally in the perturbation theory of \eqref{eq:UV1} when $h\gg \Lambda$: $\Lambda_\mathrm{n}$ is the scale where the perturbative interactions between light neutral-fermion excitations blow up, while $\Lambda_\mathrm{c}$ is the scale where interactions between light charged-fermion excitations close to the Fermi surface blow up.
More generally, we demonstrate that the perturbative expansion around different points in momentum space can hide subtle dynamically generated scales related to renormalon effects whenever large logs develop.

To further unveil the physics behind those scales, we turn to large $N$, generalizing the methods employed in \cite{Schon:2000he, Thies:2003kk, Schnetz:2004vr}. 
In the regime $h\gg \Lambda$ the Hubbard--Stratonovich field $\sigma(x) = -g^2 \sum_{j=1}^N \bar\psi^j(x) \, \psi^j(x)$ forms an inhomogeneous time-independent crystal-like condensate, which at leading order is
\begin{equation}
\label{eq:sigma-osc}
\bigl\langle  \sigma(x) \bigr\rangle \approx \Lambda_\mathrm{n} +  \Lambda_\mathrm{c} \sin(2 h x^1) \,.
\end{equation}
Here $\Lambda_\mathrm{n}$ and $\Lambda_\mathrm{c}$ govern the amplitudes of the constant part and of spatial oscillations. On the other hand, the very same scales also control the dispersion relations of light modes around (\ref{eq:sigma-osc}), with $\Lambda_\mathrm{n}$ equal to the mass gap of the neutral fermions and $\frac{1}{2}\Lambda_\mathrm{c}$ equal to the mass gap of the charged fermions.

We show that inhomogeneous condensates also appear at finite $h$, forming crystals of $a$-fermion bound states. As $h$ decreases the lattice spacing grows, until at some critical value the condensate describes a single bound state, as in \cite{Dashen:1975xh}. Such inhomogeneous condensates were already discovered for $a=N$ (also with temperature $T$), where they form out of kink-antikink bound states. The phase diagram was shown to include a crystal phase with a spatially modulated order parameter at low $T$ and sufficiently large $h$ \cite{Schon:2000he, Thies:2003kk, Schnetz:2004vr} (see also \cite{Narayanan:2020uqt, Lenz:2020bxk, Koenigstein:2021llr} for numerical lattice results at finite $N$, and \cite{Melin:2024oee} for a more recent analysis at $T=0$ using Bethe ansatz techniques).
Inhomogeneous phases were also discovered in related models \cite{Basar:2009fg, Ciccone:2022zkg, Ciccone:2023pdk}.

Finally, we use integrability to derive \eqref{eq:ts-12} at finite $N$. The Bethe ansatz makes use of the known exact spectrum of the theory, which includes bound states of $r$ fermions in antisymmetric representations of $O(2N)$ with mass
\begin{equation}
\label{eq:spectrum}
m_r = m \sin\bigl( \tfrac{\pi r}{2N-2} \bigr) \big/ \sin \bigl( \tfrac{\pi}{2N-2} \bigr) \,,\;\;\; 1\leq r \leq N-2 \,,
\end{equation}
as well as of the $S$-matrix between them. We are able to confirm that $\Lambda_\mathrm{n}$ and $\Lambda_\mathrm{c}$ govern the mass gap of the neutral and charged fermions also at finite $N$, in total agreement with the semiclassical analysis.
Besides, we observe that in this phase the GN model exhibits a \emph{massless} mode for all values of $a$, and is thus gapless. At infinite $N$ this mode is the Goldstone boson for broken translations (the phonon) in the crystal phase, as discussed in \cite{Melin:2024oee} in the case $a=N$.
At finite $N$, quantum corrections are expected to melt the crystal, as strict order is forbidden \cite{Mermin1966, Coleman1973}.
Yet, gapless excitations persist and give rise to a quasi-long-range ordered phase.

More details on the spectrum, the phase diagram, and further results at finite $N$ will be presented in a forthcoming paper \cite{long-paper}.

\medskip

\paragraph{Perturbation theory.} 
Dynamically generated scales arise where perturbation theory breaks down, which can be signaled by the appearance of large logs. At large momenta, such logs are canceled by the ultraviolet (UV) 1-loop $\beta$-function of $g^2$, and the theory is asymptotically free. This UV $\beta$-function seems to imply a breakdown at the scale $\Lambda$, which is indeed the case when $h=0$. But with nonzero $h$, other scales might emerge as we approach some finite momenta, as we will show.
At nonvanishing chemical potential the fermion propagator is
\begin{equation}
\bigl\langle \psi^i_{\alpha}(p) \, \bar\psi^j_{\beta}(-p) \bigr\rangle = \delta_{ij} \, \biggl( \frac{i}{\slashed{p} + h_i\gamma^0} \biggr)\!\rule[-0.8em]{0pt}{0cm}_{\alpha\beta} \;,
\end{equation}
where $\alpha, \beta$ are spinor indices, while $h_i = h$ for charged fermions, $h_i=0$ for the neutral ones. The position-space propagator is time ordered, dictating that the contour for $p^0$ in loop integrals can be Wick rotated to Euclidean momentum $p^2 = -ip^0$ without intersecting any pole. When $h_i = 0$, this is the Feynman $i\epsilon$ prescription. The 1-loop correction to the propagator is independent of the external momentum, and does not affect the running of $g$. 

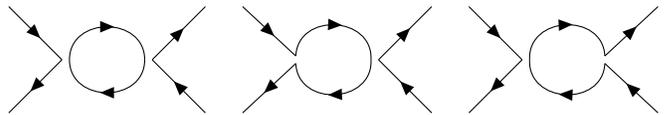
\begin{figure}
\centering
\begin{tikzpicture}
\begin{feynman}[small]
	\vertex (v1l);
	\vertex[right=0.1 cm of v1l] (v1r);
	\vertex[right=of v1r] (v2l);
	\vertex[right=0.1 cm of v2l] (v2r);
	\vertex[above left=of v1l] (i1);
	\vertex[below left=of v1l] (i2);
	\vertex[below right=of v2r] (i3);
	\vertex[above right=of v2r] (i4);
	\diagram*{(i1) --[fermion] (v1l) --[fermion] (i2), (i3) --[fermion] (v2r) --[fermion] (i4), (v1r) -- [fermion, half left] (v2l) --[fermion, half left] (v1r)};
\end{feynman}  
\end{tikzpicture}
\hfill
\begin{tikzpicture}
\begin{feynman}[small]
	\vertex (v1);
	\vertex [above = 0.05cm of v1](v1t);
	\vertex[below=0.05cm of v1] (v1b);
	\vertex[right=of v1] (v2l);
	\vertex[right=0.1 cm of v2l] (v2r);
	\vertex[above left=of v1] (i1);
	\vertex[below left=of v1] (i2);
	\vertex[below right=of v2r] (i3);
	\vertex[above right=of v2r] (i4);
	\diagram*{(i1) --[fermion] (v1t) --[fermion, half left] (v2l) --[fermion, half left] (v1b) --[fermion] (i2), (i3) --[fermion] (v2r) --[fermion] (i4)};
\end{feynman}  
\end{tikzpicture}
\hfill
\begin{tikzpicture}
\begin{feynman}[small]
	\vertex (v1);
	\vertex [left=0.05cm of v1](v1l);
	\vertex[right=0.05cm of v1] (v1r);
	\vertex[right=of v1r] (v2);
	\vertex[above=0.05cm of v2] (v2t);
	\vertex[below=0.05 cm of v2] (v2b);
	\vertex[above left=of v1l] (i1);
	\vertex[below left=of v1l] (i2);
	\vertex[above right=of v2] (i4);
	\vertex[below right=of v2] (i3);
	\diagram*{(i1) --[fermion] (v1l) --[fermion] (i2), (i3) --[fermion] (v2b) --[fermion, half left] (v1r) --[fermion, half left] (v2t) --[fermion] (i4)};
\end{feynman}  
\end{tikzpicture}
\caption{\label{fig:1-loop-correlator}%
One-loop diagrams that contribute to $\Gamma$. Vertices are slightly separated to track spinor contractions.}
\end{figure}

Let $\Gamma \equiv \bigl\langle \bar\psi^i_\alpha(\frac{q}{2}) \, \psi^i_\beta(\frac{q}{2}) \, \bar\psi^j_\gamma(-\frac{q}{2}) \, \psi^j_\delta(-\frac{q}{2}) \bigr\rangle$ be the (amputated) four-point correlator with all incoming momenta. For brevity, we assume that $i\neq j$ and both species are charged or both neutral. At 1-loop there are three diagrams that contribute to the amplitude
\footnote{There are two more diagrams that contribute to general four-point correlators, but they vanish for our particular choice of momenta.},
see Fig.~\ref{fig:1-loop-correlator}. 
All fermion loops in these diagrams contribute through an integral of the form
\begin{equation}
\int \! \frac{d^2p}{(2\pi)^2} \;\; \frac{i}{\slashed{p}+h_i\gamma^0} \;\; \frac{i}{\slashed{p} - \slashed{q} + h_i\gamma^0}
\end{equation}
which we regulate using a cutoff $M$ on the spatial momentum only: ${\int\! d^2p \to \int^{M}_{-M} dp^1 \int \! dp^0}$. After Wick rotation we can close the contour for $p^2$ and pick up the appropriate poles, and then perform the $p^1$ integration. Including the tree-level contribution and a counterterm $\delta_g$ for the $(\bar\psi \psi)^2$ interaction, we obtain
\begin{multline}
\Gamma = - g^4 A(h_i) \bigl(\gamma^*_{\alpha\beta} \epsilon_{\gamma\delta} + \epsilon_{\alpha\beta} \gamma^*_{\gamma\delta} \bigr) + \epsilon_{\alpha\beta} \epsilon_{\gamma\delta} \Bigl[ ig^2 + {} \\
+ i\delta_g - 2g^4\bigl((N-a)B(0) + aB(h) - B(h_i)\bigr) \Bigr] ,
\end{multline}
where $\gamma^* = \gamma^1\gamma^0 = \begin{psmallmatrix}1 & 0 \\ 0 & -1 \end{psmallmatrix}$ is the chiral gamma matrix,  $\epsilon$ the antisymmetric tensor, and 
\begin{equation}
A(h) = \frac{i}{4\pi}\log \biggl( \frac{\tilde q_+}{\tilde q_-} \biggr), \quad
B(h) = \frac{i}{4\pi} \log \biggl( \frac{\tilde q_+\tilde q_-}{4M^2} \biggr) ,
\end{equation}
with $\tilde q_\pm^2 \equiv (2h \pm q_1)^2 + q_2^2$.
There are three points where the amplitude could blow up: when $q \equiv (q_1, q_2) \to (0,0)$ due to light neutral fermions running in the loop, and when $q \to q_F^\pm \equiv (\pm 2h, 0)$ due to light charged fermions running in the loop
\footnote{$\Gamma$ develops large logs also at $q \to \infty$: this leads to the UV $\beta$-function of \cite{Gross:1974jv} and to $\Lambda$ in (\ref{Lambda}).}.

Starting from weak coupling in the UV, as we approach those points, we determine the scales that control at what distance perturbation theory breaks down.

In order to approach the first point, $q \to 0$, we fix the amplitude involving two neutral species $r$ and $s$, and parametrize $q = \mu \, \hat n$ using a unit vector $\hat n = (n_1, n_2)$. The counterterm $\delta_g$ is then set by demanding that
\begin{equation}
\Gamma^{rs}_{\alpha\beta\gamma\delta}(\mu \hat n) = i g^2 \, \epsilon_{\alpha\beta} \epsilon_{\gamma\delta} \,.
\end{equation}
At scales $\mu \ll h$ we find $\delta_g = \frac{N-a-1}{2\pi} \, g^4 \log \bigl( \mu^2 / 4 M^2 \bigr)$.
The $\beta$-function $\beta = \partial g^2 / \partial \log\mu$ can now be computed as usual using the Callan--Symanzik equation, which after standard manipulations gives, at 1-loop level: 
\begin{equation}
\beta = - \mu \, \frac{\partial \delta_g}{\partial \mu} = - \frac{N-a-1}{\pi} \, g^4 \qquad (\text{as  } q\to 0) \,.
\end{equation}
Consequently, the theory becomes strongly coupled around the scale $\Lambda_\mathrm{n} \ll \Lambda$ defined in \eqref{eq:lambda-12}.
Note that this is the same result we would have gotten by adding a mass term of order $h$ for a number $a$ of fermions in the UV Lagrangian \eqref{eq:UV1} in the absence of a chemical potential. 

In order to approach the Fermi surface, say at $q \to q_F^+$, we choose to define the coupling using an amplitude that involves two charged species, $m$ and $n$, and parametrize $q = q_F^+ + \mu \, \hat n$. The counterterm can only be used to fix a single spinor component of the amplitude, and we choose the one that determines the particle-hole scattering near the Fermi surface:
\begin{equation}
\Gamma_{-++-}^{mn} \bigl( q_F^+ + \mu\hat n \bigr) = ig^2 \, \epsilon_{-+}\epsilon_{+-} = -ig^2 \,.
\end{equation}
Again we set the counterterm and find the 1-loop $\beta$-function. The latter in the regime $\mu \ll h$ simplifies to
\begin{equation}
\beta = - \frac{a}{2\pi} \, g^4 \qquad (\text{as  }q\to q_F^+) \,,
\end{equation}
leading to the scale $\Lambda_\mathrm{c} \ll \Lambda$ in \eqref{eq:lambda-12}.

To make sure that $\Lambda_\mathrm{c}$ is indeed the correct dynamically generated scale, we need to show that perturbation theory does not break down before at a higher scale, \ie, verify that all other components of the amplitude are under control as long as $g \ll 1$
\footnote{Without this test, one would be ambiguously led to other slightly different scales depending on the component of $\Gamma$ one starts with.}.
For example, by resumming a family of diagrams where we concatenate the loops in Fig.~\ref{fig:1-loop-correlator} one after the other, we find
\begin{align}
\Gamma^{mn}_{+--+} &= -ig^2 \, \Bigl( 1 + 4ig^2 A(q) + \bigl( 4ig^2 A(q) \bigr)^2 + \ldots \Bigr) \nonumber \\
&= \frac{-ig^2 }{ 1 + \frac{g^2}{2\pi} \log\Bigl( \frac{(4h+\mu \hat n_1)^2 \, + \, \mu \hat n_2^2}{\mu^2} \Bigr) } \,,
\end{align}
which remains small as long as $g \ll 1$. The same holds for all other components, confirming our result.

Summarizing, perturbation theory produces two new scales: $\Lambda_\mathrm{n}$ associated with neutral fermions; $\Lambda_\mathrm{c}$ associated with charged fermions, which appears when the momentum is near $2h$ and thus it hints at spatial oscillations on a scale of $1/(2h)$.

\medskip

\paragraph{Large $N$.}
A large $N$ analysis reveals the physics behind the two nonperturbative scales: a crystal phase with inhomogeneous condensates. To show it, we introduce a Hubbard--Stratonovich scalar field $\sigma(x)$ (as in \cite{Gross:1974jv}) and integrate the fermions out. After Wick rotation, we obtain the Euclidean action
\begin{equation}
\label{eq:action-sigma}
\!\frac{S_\text{E}}N = - y \Tr \log \mathcal{D}_h - (1{-\,}y) \Tr \log \mathcal{D}_0 + \!\int\! d^2x \, \frac{\sigma^2}{2\lambda} \,,
\end{equation}
where $\mathcal{D}_h = \gamma^0 \bigl( \slashed{\partial}_\text{E} + \sigma - \gamma^0 h \bigr)$ and $\lambda= N g^2$ is the 't~Hooft coupling. 
We take $N\to \infty$ and $a\to \infty$ while keeping the ratio $y=a/N$ fixed.
The resulting equation of motion for $\sigma$ can be written in terms of Green's functions (\ie, the fermion propagators):
\begin{equation}
\label{eq:EoM-sigma}
\!\! \frac{\sigma(x)}{\lambda} = y \tr \bigl( \mathcal{D}_h^{-1}(x,x) \gamma^0 \bigr) + (1 {-\,} y) \tr \bigl( \mathcal{D}_0^{-1}(x,x) \gamma^0 \bigr)
\end{equation}
where the trace is now over the spinor indices only. When $h=0$, the solution is given by the constant $\sigma = \Lambda$. With the chemical potential turned on, one finds a number of homogeneous solutions that depends on $h$. However, like in \cite{Koenigstein:2021llr}, at sufficiently high $h$ the leading homogeneous solution becomes unstable and an inhomogeneous solution is preferred instead \cite{long-paper}.

\begin{figure}
\centering
\includegraphics[width=0.85\linewidth]{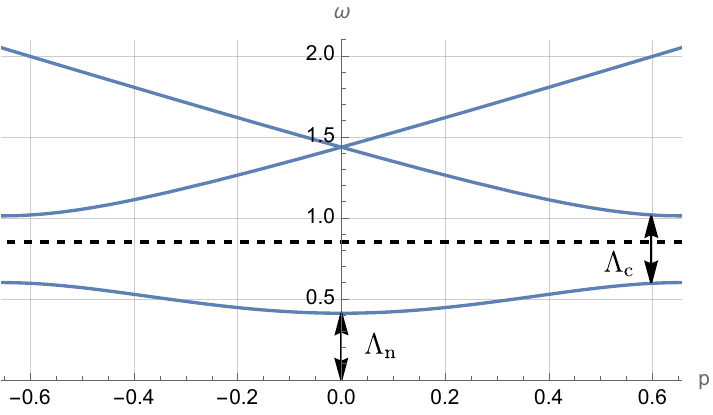}
\caption{\label{fig:disp-analytical}%
The two-band structure of $\omega(p)$ in the Brillouin zone obtained from \eqref{eq:disp-rel} for $y=2/3$ and $h= 0.85 \, \Lambda$ (dashed line).}
\end{figure}

To find the inhomogeneous solution, like in \cite{Thies:2005wv, Thies:2003br, Schnetz:2005ih}, we use the time-independent ansatz
\begin{equation}
\sigma(x) = A \, \biggl( \mathfrak{m} \sn(b) \sn(\tilde{x}) \sn(\tilde{x}+b) + \frac{\cn(b) \dn(b)}{\sn(b)} \biggr) ,
\end{equation}
where $\tilde{x}=Ax^1$ is a rescaled spatial component, and $\sn$, $\cn$, and $\dn$ are the Jacobi elliptic functions with implicit parameter $\mathfrak{m}$. The equations of motion are solved by diagonalizing the operator $\mathcal{D}_h$ with this ansatz as in \cite{Schnetz:2005ih}. This fixes two of the parameters $\{A,\mathfrak{m},b\}$ in terms of the third one, and choosing the dominant saddle fixes the latter in terms of $h$. 
The general equations, their detailed derivation and solutions, and a comparison of the free energies are deferred to \cite{long-paper}. In the high-density regime $h \gg \Lambda$, the leading-order solution is
\begin{equation}
A \approx h \,, \;\;\;
\mathfrak{m} \approx 4 \, \Bigl( \frac{\Lambda}{2h} \Bigr)^{\!\frac{2}{y}} , \;\;\;
b \approx K(\mathfrak{m}) - 2 \, \Bigl( \frac{\Lambda}{2h} \Bigr)^{\!\frac{1}{1-y}} \!,
\end{equation}
where $K$ is the complete elliptic integral of the first kind.

The eigenvalue densities of $\mathcal{D}_h$ and $\mathcal{D}_0$ can be viewed as the dispersion relations for charged and neutral fermion excitations, respectively, on top of the background $\sigma$:
\begin{equation}
\label{eq:disp-rel}
\frac{dp}{d\omega} = \frac{\tilde\omega \, \left\lvert \tilde\omega^2-s+1-E(\mathfrak{m})/K(\mathfrak{m}) \right\rvert }{ \sqrt{ ( \tilde\omega^2 - s + 1 ) ( \tilde\omega^2 - s + \mathfrak{m} )  ( \tilde\omega^2 - s ) }} \;,
\end{equation}
with $s={\rm sn}^{-2}(b)$, $E$ the complete elliptic integral of the second kind, and $\tilde\omega = \omega/A$. An example of dispersion relation is given in Fig.~\ref{fig:disp-analytical}, and it mirrors the ones emerging from Peierls instability in 1d conductors.
This spectrum has three gaps: one at the bottom of the lower band between $0$ and $A\sqrt{s-1}$, corresponding to the gap for neutral fermions; one from $A \sqrt{s - \mathfrak{m}}$ to the Fermi surface at $h$, corresponding to the gap for charged holes; one from $h$ to $A\sqrt{s}$, corresponding to the gap for charged fermions. In the large $h$ limit the latter two become identical, which is evidence for an effective relativistic dispersion near the Fermi surface.
The gaps match with the large $N$ limit of the scales in \eqref{eq:lambda-12}: 
\begin{equation}
\label{exact scales}
\begin{aligned}
A\sqrt{s-1} \,\approx\, (2h)^{1-\frac{1}{1-y}} \, \Lambda^{\frac{1}{1-y}} &\approx \Lambda_\mathrm{n} \,, \\
A \bigl( \sqrt{s} - \sqrt{s-\mathfrak{m}} \bigr) \,\approx\, (2 h)^{1-\frac{2}{y}} \, \Lambda^{\frac{2}{y}} &\approx \Lambda_\mathrm{c} \,.
\end{aligned}
\end{equation}

The profile of $\sigma(x)$ in the large $h$ limit simplifies to \eqref{eq:sigma-osc}. Thus we identify the two scales with parameters of the crystal phase:
$\Lambda_\mathrm{n}$ replaces $\Lambda$ as the constant part of the condensate, while $\Lambda_\mathrm{c}$ modulates the periodic oscillations with frequency $2h$ as anticipated by the perturbative result near the Fermi surface.  In the limit $y\rightarrow 1$, the scale $\Lambda_\mathrm{c}$ becomes the band gap for fermions identified in \cite{Melin:2024oee}.

\begin{figure}
\centering
\includegraphics[width=\linewidth]{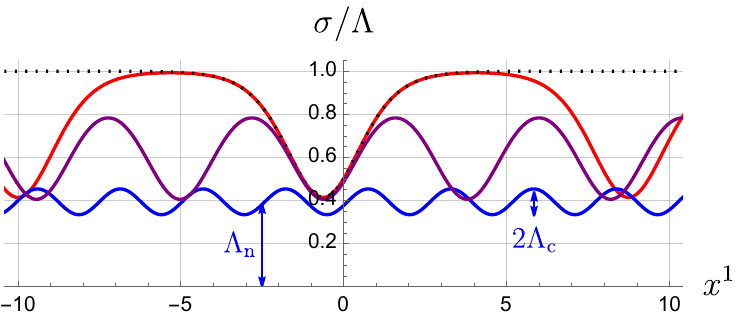}
\caption{\label{fig:sigma-profile}%
The profile of $\sigma(x)$ for $y=1/2$, for very low density ($h=1.0001 \, h_*$, in red), medium density ($h=0.95 \, \Lambda$, in purple) and high density ($h= 1.3 \, \Lambda$, in blue). The dotted line is the single bound state from \cite{Dashen:1975xh}. Note how the high-density approximation \eqref{eq:sigma-osc} extends up to values of $h\gtrsim \Lambda$.} 
\end{figure}

As we decrease $h$, the period of the lattice grows and the density lessens. At the critical value 
\begin{equation}
h_* = \frac{2 \sin \bigl(  \frac{\pi y}{2} \bigr)} {\pi y}  \, \Lambda \,,
\end{equation}
the parameters of the solution are 
\begin{equation}
\mathfrak{m}=1 \,,\qquad A = \Lambda \tanh(b) = \Lambda \sin \bigl(  \tfrac{\pi y}{2} \bigr) \,,
\end{equation}
and the condensate $\sigma(x)$ becomes a single localized bound state of $a$ particles, exactly as described in \cite{Dashen:1975xh} (see \cite{long-paper} for more details). The critical value $h_*$ corresponds to the mass-to-charge ratio derived from \eqref{eq:spectrum} (at large $N$, $m=\Lambda$), and marks the second order
\footnote{The phase transition is due to the large $N$ limit. In any finite volume $L$ there is a first-order phase transition which becomes smoother as $L$ grows, leading to a second-order transition in infinite volume.}
phase transition to the homogeneous phase $\sigma = \Lambda$. This confirms that the vacuum is a crystal of $a$\hspace{0.1em}-particle bound states, which plays an important role in the integrability analysis.
We compare the low-density clearly separated crystal with the high-density sinusoidal curve in Fig.~\ref{fig:sigma-profile}.

\medskip

\paragraph{Integrability.}
Bethe ansatz techniques can be used to identify the two scales even for \textit{finite} $N$. 
Since the particle with the smallest mass-to-charge ratio is the rank-$a$ bound state with polarization aligned with the charge, at $T=0$ these particles start to populate the vacuum as soon as the chemical potential is larger than that ratio. The ground state is characterized by the equation \cite{Polyakov:1983tt}
\begin{equation}
\label{eq:TBA}
\epsilon(\theta) - \!\int_{-B}^B \! K_a(\zeta - \theta) \, \epsilon(\zeta) \, d\zeta = a h - m_a \cosh(\theta) \,.
\end{equation}
Here $\epsilon$ is the energy distribution in rapidity space, while $B$ is the Fermi rapidity that satisfies the boundary condition $\epsilon(\pm B) = 0$. 
Lastly, $K_a(\theta)= \frac{1}{2\pi i} \, \partial_\theta \log S_a(\theta)$, with $S_a$ the $S$-matrix of two bound states scattering, which is known exactly and is found, \eg, in \cite{Fendley:2001hc}. The free energy is calculated directly:
\begin{equation}
\Delta F(h) = - \frac{m_a}{2\pi} \int_{-B}^B \epsilon(\theta) \cosh(\theta) \, d\theta \,,
\end{equation}
valid for $h$ above the condensation threshold.

At large $h$ we have ${B\propto \log(h/\Lambda)}$. This allows an expansion in powers of $1/B$ through Volin's method \cite{Volin:2009wr, Marino:2019eym} supplemented, through Wiener--Hopf analysis, with nonperturbative terms proportional to powers of $\Lambda/h$, leading to the full transseries of the free energy \cite{Marino:2021dzn, Bajnok:2022xgx} (see \cite{Reis:2022tni} for a review). A key result of this analysis is that nonperturbative effects in the free energy are proportional to $h^2 (\Lambda/h)^{2\xi_k}$ with the $\xi_k$'s given by the positive real poles of $G(-i \xi)/G(i \xi)$. Here $G$ is the Wiener--Hopf decomposition of the Fourier transform of the kernel: an analytic function in the upper half-plane such that $1-K(\omega) = \bigl( G(\omega) \, G(-\omega) \bigr){}^{-1} $. For the integral equation \eqref{eq:TBA}, it is given by
\begin{equation}
\label{Gplus-a}
\!\! G(\omega) = e^{\tfrac{ i\omega \mathfrak{b} + \ell_{-i\omega}}{2N-2} }
\frac{ \Gamma \bigl( 1 - \tfrac{i a\omega}{2N-2} \bigr) \, \Gamma \bigl( \frac{1}{2} - \tfrac{ (N-a-1) i\omega }{ 2N-2}  \bigr) }{ \sqrt{a} \;
\Gamma \bigl( \frac{1}{2} - \frac{i \omega }{2} \bigr) \, \Gamma \bigl( 1 - \tfrac{i\omega }{ 2N-2} \bigr)},
\end{equation}
where $\ell_x = x \log x$ and 
\begin{equation}
\mathfrak{b}= 1+ \log (2N-2)+ \ell_a + \ell_{N-a-1}-\ell_{N-1} \,.
\end{equation}

Inspecting \eqref{Gplus-a}, one finds two families of poles:
\begin{equation}
\label{res-a}
\xi_k =  (2 k - 1) \, \frac{N-1}{N-a-1} \,,\qquad \xi'_k = 2 k \, \frac{N-1}{a} \,,
\end{equation}
with $k\in\mathbb{N}$. These poles lead to the nonperturbative effects in \eqref{eq:ts-12}. 
The exact coefficients $a_{k,p}^{[m,n]}$ and $c_0$ therein can be determined analytically following a generalization of the procedure outlined in \cite{Marino:2021dzn}. While at certain values of $N$ and $a$ some $\xi_k$ and $\xi'_{k'}$ might overlap creating a double pole, in the cases where this does not happen we have $k_{m,n} = p_{m,n} = 0$ and \eqref{eq:ts-12} simplifies. For the specific case $a=1$, $\Lambda_\mathrm{c}$ does not appear because the residues associated with $\xi'_k$ vanish, and $(\Lambda_\mathrm{n})^k$ are the ``new renormalons'' in \cite{Marino:2021dzn}. It would be interesting to understand whether \eqref{eq:ts-12} can be reproduced with condensates as in \cite{Marino:2024uco} supplemented with \eqref{eq:sigma-osc}.

\begin{figure}
\centering
\includegraphics[width=\linewidth]{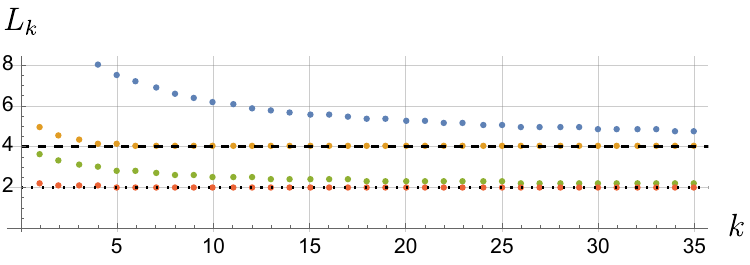}
\caption{\label{fig:exp-plots}%
The sequence $L_k$ in \eqref{eq:seq-Lk} for $N=5$ and $a=2$, for charged fermions/holes (in blue, with Richardson transform \cite{Bender1978} in orange) and neutral fermions (in green, with Richardson transform in red). The dashed and dotted lines correspond to $(2N-2)/a$ and $(N-1)/(N-a-1)$, respectively.}
\end{figure}

From the Bethe ansatz we can also extract the mass gaps at finite $N$ by considering probe particles on top of the ground state.
A probe that creates a single $b$\hspace{0.1em}-particle bound state has an energy distribution $\bar\epsilon_b$ that satisfies
\begin{equation}
\label{eq:TBA-probe}
\bar\epsilon_b(\theta) - \!\int_{-B}^B \!\! K_{ab}(\zeta - \theta) \, \epsilon(\zeta) \, d\zeta = q_{ab} h - m_b \cosh(\theta) \,,
\end{equation}
where $q_{ab} = \min(a,b)$ is the charge carried by the particle under the chemical potential and $K_{ab}(\theta)$ is obtained from the scattering of an $a$ and a $b$ bound states (see \cite{Fendley:2001hc}). A charged fermion probe corresponds to simply adding a fermion ($b = 1$), while a charged hole removes a particle from a background bound state ($b = a - 1$). Lastly, a neutral fermion binds to one of the background baryons forming a larger bound state ($b = a + 1$). The value of $\bar\epsilon_b$ at zero rapidity is proportional to the gap. We test its scaling with $h$ by evaluating sequences such as
\begin{equation}
\label{eq:seq-Lk}
L_k = 1-\tfrac{8}{k}\log \bigl( m^{-1} \, \bar\epsilon_b ( \theta=0; h = e^{k/8} m ) \bigr) \,.
\end{equation}
As exemplified in Fig.~\ref{fig:exp-plots}, we numerically find that for the neutral fermions the sequence converges (up to $1/k$ corrections) to ${(N-1)/(N-a-1)}$, while for charged fermions/holes it converges to ${(2N-2)/a}$, reproducing the scales $\Lambda_\mathrm{n}$ and $\Lambda_\mathrm{c}$, respectively.

\begin{figure}
\centering
\includegraphics[width=\linewidth]{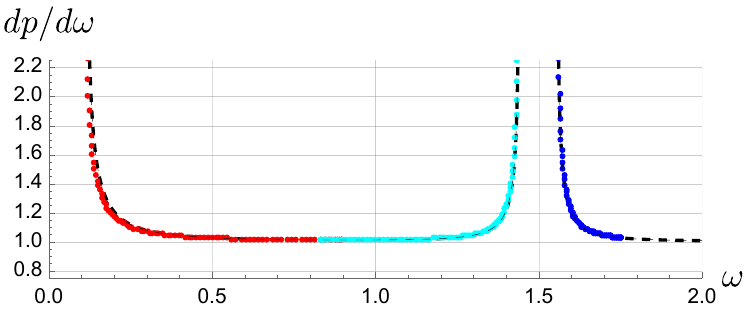}
\caption{\label{fig:disp-num}%
Plot of $dp/d\omega$ for $N=102$, $a=68$, $h=1.5 \, \Lambda$.  The points are the numerical solution of \eqref{eq:TBA-probe} for charged fermions (in blue), charged holes (in cyan), and neutral fermions (in red). The dashed line is  \eqref{eq:disp-rel}.}
\end{figure}

One can also extract the dispersion relations for each type of probe. For neutral fermions and charged fermions/holes, at large $N$ they reproduce the exact result \eqref{eq:disp-rel}, as shown in Fig.~\ref{fig:disp-num}. Near the Fermi surface, both at large and finite $N$, we find from \eqref{eq:TBA} that excitations of the background itself are \emph{gapless}, similarly to what was found in the case $a=N$ in \cite{Melin:2024oee}.

\medskip

In conclusion, the Gross--Neveu model is a perfect arena where connections between perturbative and nonperturbative effects, including renormalons and condensates, can be explicitly studied, thanks to large $N$ and integrability techniques. In particular, we showcased how the analysis of large logs in perturbation theory in momentum space can reveal new dynamically generated scales. Especially intriguing are the crystal-like phases arising when chemical potentials are turned on. They display a gapless mode that at infinite $N$ is interpreted as the Goldstone mode for broken translations (phonon) and that persists at finite $N$. Similar phenomena are likely to occur in other 2d integrable models, such as the principal chiral model, and perhaps even in four-dimensional QCD (see, \eg, Section~7 of \cite{Fukushima:2010bq}).

\medskip

\begin{acknowledgments}
\textit{Acknowledgments} --- We thank Marcos Mari\~no and Konstantin Zarembo for useful discussions and correspondence.
This work was supported by the ERC-COG Grant NP-QFT No. 864583,
the MUR-FARE2020 Grant No. R20E8NR3HX,
and the MUR-PRIN2022 Grant No. 2022NY2MXY.
We are also supported by the INFN ``Iniziativa Specifica"  GAST and  ST\&FI.
\end{acknowledgments}

\bibliography{renormalons_prl.bib}

\end{document}